\documentstyle[a4,12pt,amssymb]{article}
\input epsf
\begin{document}
\baselineskip=20pt

\title{Integrals of motion and the shape of the attractor
for the Lorenz model}
 
\author{H.Giacomini \footnote{email: giacomini@univ-tours.fr} $\,$and
 S.Neukirch \footnote{email: seb@celfi.phys.univ-tours.fr} \\
C.N.R.S. Laboratoire de Math\'ematiques et de Physique Th\'eorique\\
Facult\'e des Sciences et Techniques\\
Parc de Grandmont\\
Universit\'e de Tours\\
37200 Tours\\
France\\}
\date{}
\maketitle

\vspace{1cm}

{\small
\begin{center}
\section*{Abstract}
\end{center}
In this paper, we consider three-dimensional dynamical
systems, as for example the Lorenz model. For these systems,
we introduce a method for obtaining families of
two-dimensional surfaces such that trajectories cross
each surface of the family in the same direction.
For obtaining these surfaces, we are guided by the
integrals of motion that exist for particular values
of the parameters of the system. Nonetheless families of
surfaces are obtained for arbitrary values of these
parameters.
Only a bounded region of the phase space is not filled by
these surfaces. The global attractor of the system must be
contained in this region.
In this way, we obtain information on the shape and location
of the global attractor. These results are more restrictive
than similar bounds that have been recently found by the
method of Lyapunov functions.\\
\\
\\
\\

{\bf Keywords} : Lorenz model/ Chaotic Dynamics/ Integrals of motion.

{\bf PACS numbers} : 05.45.+b / 02.30.Hq
}
\clearpage
The Lorenz equations (\ref{lo equa}) are one of the
classic models of nonlinear dynamics and chaos. These
equations were originally derived in a modal truncation
of the Boussinesq equations for thermal convection. They
read as follows~:
\begin{eqnarray}
\dot{x} & = & \sigma(y-x) \nonumber \\
\dot{y} & = & rx-y-xz  \label{lo equa}\\
\dot{z} & = &xy-bz  \nonumber 
\end{eqnarray}
with $\sigma , b, r \geq 0 $.
There $\sigma$ corresponds to the Prandtl number, $b$ is
a geometric parameter and $r$ is the Rayleigh number in
units of the critical Rayleigh number.

These equations describe a dissipative dynamical system
for all values of $r$, $\sigma$ and $b$ because the
divergence of the flow field is always negative. 
Hence 3-dimensional volumes in the phase space contract
to zero at a uniform exponential rate and the system's
attractor is necessarily of dimension less than three.
This model has become a classic in the area of nonlinear
dynamics. Its importance is not that it quantitatively
describes the hydrodynamics motion, but rather that it
illustrates how a simple model can produce very rich and
varied form of dynamics, depending on the value of a
parameter in the equations \cite{sparrow}.

In this paper, we are interested in the approximated
location in the phase space of the global attractor of
the system, which contains all dynamics evolving from
all initials conditions. The global attractor is the set
of points in phase space that can be arrived at from some
initial condition at an arbitrary long time in the past.
The two fundamental properties of global attractors
are \cite{doering navier}~:
\begin{itemize}
\item it is invariant under the evolution.
\item the distance of any solution from it vanishes as
$t \rightarrow + \infty $.
\end{itemize}
The last property is simply interpreted thus~: if the
solution starts initially outside the global attractor,
then it is attracted into it as $t \rightarrow + \infty $
and once inside it cannot escape. If it starts inside then
it stays inside. 

The global attractor contains all the
asymptotic motion for the dynamical system. It is common
to talk of {\it multiple attractors} for a dynamical system,
and each of them  may in its own right be considered as the
attractor for initial conditions within its own bassin
of attraction. The notion of global attractor
corresponds to the union of all possible such dynamically
invariant attracting sets. In particular, it contains all
possible structures such as fixed points, limit
cycles etc...

The global attractor is contained in an {\it absorbing ball} in phase
 space, and we want to obtain analytic estimates about its
geometric shape. Moreover, this enables us to find good
estimates of its Lyapunov dimension. Estimates which give
the shape of the attractor are important as they lead to
 a good upper bound on the dimension of the
Lorenz attractor \cite {doering et 
gibbon}.

Until now, approximate locations of the
Lorenz's attractor in the phase space have been obtained
by the method of Lyapunov functions \cite{lorenz,sparrow,
doering et gibbon,strogatz,jackson}. Very recently, thanks
to this method, it has been shown that the global
attractor of the Lorenz equations is contained in a
volume bounded by a sphere, a cylinder, the volume 
between two parabolic sheets, an ellipsoide and a cone
\cite{doering et gibbon}.\\
In this paper, we apply a different method for obtaining
analytic estimates for the location and shape of the
Lorenz attractor. The method is based on the determination
of families of 2-dimensional surfaces that are crossed by
the trajectories of the system only in one direction. In
the region filled by these surfaces, the dynamical
behaviour is very simple. The asymptotic {\it complex 
behaviour} must be contained in the region of the
phase space that is not occupied by these surfaces.\\

\begin{figure}[htbp]
$$
\epsfxsize=5cm
\epsfbox{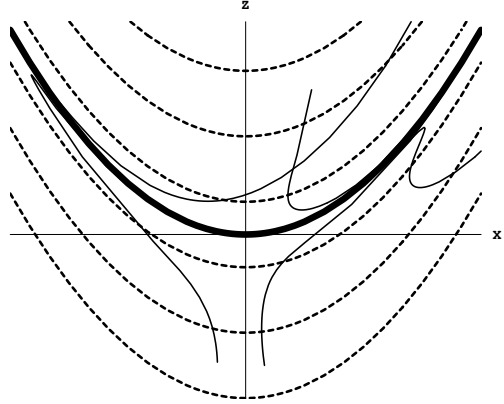}
$$
\caption{The family of surfaces $x^2-2 \sigma z=k$ for
 the case $b=2 \sigma$.
The bolded surface corresponds to $k=0$ and is the
 attracting set of the system. Some trajectories
 of the system are shown.}
\label{fig para m}
\end{figure}

For finding these families of surfaces, we will be guided
by the time-dependent integrals of motion that exist for
special values of the parameters of the system.
Integrals of motion for the Lorenz system have been
extensively studied in \cite{tabor,kus,tabor et weiss}.
The known integrals of motion are~:
\begin{itemize}
\item [{\bf a}] $I(x,y,z,t)=(x^2-2 \sigma z)  e^{2 \sigma t}$
 with  $b=2 \sigma$  and  $\sigma$  and  $r$ arbitrary.

\item [{\bf b}] $I(x,y,z,t)=(y^2+z^2) e^{2t}$
with $b=1$, $r=0$ and $\sigma$ arbitrary.

\item [{\bf c}] $I(x,y,z,t)=(-r^2x^2+y^2+z^2) e^{2t}$
with $b=1$, $\sigma=1$ and $r$ arbitrary.

\item [{\bf d}] $I(x,y,z,t)=\left( \frac{(2 \sigma-1)^2}{\sigma}
 x^2+
\sigma y^2-(4 \sigma-2)xy-\frac{1}{4 \sigma} x^4+x^2 z
\right)
e^{4 \sigma t}$ \\ with $b=6 \sigma-2$, $r=2 \sigma
-1$ and $\sigma$ arbitrary.

\item [{\bf e}] $I(x,y,z,t)=\left(
-rx^2-y^2+2xy+\frac{1}{4}x^4-x^2 z+
4(r-1)z
\right) 
e^{4t}$ \\ with $b=4$, $\sigma=1$ and $r$
arbitrary.
\end{itemize} 
For each of these integrals we have 
$\frac{dI}{dt} \equiv 0$.
Let us consider case {\bf a} and let us define the family 
of surfaces $x^2-2 \sigma z=k$, where k is an arbitrary 
constant. The scalar product between the normal vector 
$\vec{N}$ to this surface at a given point  and the 
tangent vector $\vec{T}$ to the trajectory of the Lorenz
system that goes through this point is given by~:
\begin{equation}
\vec{N} \cdot \vec{T}= (2x \vec{i} - 2 \sigma \vec{k})
\cdot \left( \sigma (y-x) \vec{i}+(rx-y-xz) \vec{j}+
(xy-bz) \vec{k} \right) =-bk \nonumber 
\end{equation}
Therefore, for a given surface (i.e. for a given value of k)
this scalar product has the same sign for all the points of
the surface. Each surface of the family is crossed
 in the same direction by the flow associated to the system.
This direction depends of the sign of the constant
k. Hence, for the case $b=2 \sigma$, the 3-d phase 
space of the
Lorenz system is filled by two families of surfaces, the
families associated to positive and negative values of k.
 The scalar product 
$\vec{N} \cdot \vec{T}$ is positive (resp. negative) for
 negative (resp. positive) values of k. It is clear that
the surface corresponding to k=0 plays a very special role.
All the trajectories of the system are attracted by this
surface. On this surface, the scalar product
$\vec{N} \cdot \vec{T}$ is zero. This surface is an
invariant manifold of the system, as can be seen 
in fig. \ref{fig para m}.

It is clear that the existence of these 
families of surfaces gives a lot of information about the
dynamics of the system. The behaviour of trajectories is
extremely simple in all the phase space with the
exception of the invariant surface $x^2-2 \sigma z=0$.
This surface contains the global attractor of the system for
the case $b=2 \sigma$. Here, the global attractor is contained 
in a two-dimensional surface, as for the five cases 
{\bf a}, {\bf b}, {\bf c}, {\bf d} and {\bf e}, that is when an
 integral of motion
exists. The family of surfaces derivated above enables us to
characterize in a simple way this global attractor. The
determination of this family of surfaces follows immediatly
from the existence of the integral of motion  {\bf a}, when
$b=2 \sigma$. Now, the natural question is~: if 
$b \not= 2 \sigma$, is it possible to find similar
families of surfaces that the flow crosses in the
same direction at each point of the surface ? - in the 
following, we will call this type of surface
{\it semipermeable} -
In this case, we do not have at our disposal an integral of
motion, and these surfaces can not fill the phase space
because, in the general case, the global attractor is not
contained in a two-dimensional set.\\

In order to find {\it semipermeable} surfaces in the general case 
(when integrals of motion do not exist),
we will procede as follows~:\\
For the case of the integral of motion {\bf a}, 
we first propose, when $b \neq 2 \sigma$, a surface of the 
same mathematical form as the integral of motion {\bf a}, but
with arbitrary coefficients~:
\begin{equation}
S=a_1 x^2+a_2z+a_3=0 \label{para}
\end{equation}
The scalar product $\vec{N} \cdot \vec{T}$ is now~:
\begin{eqnarray}
\vec{N} \cdot \vec{T} &=& 2 a_1 x \dot{x}+a_2 \dot{z} 
\nonumber \\
&=&(2a_1 \sigma +a_2)xy-2 a_1 \sigma x^2 -a_2 b z 
\label{ps para} 
\end{eqnarray}
If we calculate this scalar product on the surface S, for the general
case $b \neq 2 \sigma$, we obtain ~:
\begin{equation}
\vec{N} \cdot \vec{T}/S=(2a_1 \sigma+a_2)xy+(b a_1-2 \sigma
a_1)x^2+b a_3 \label{ps para on}
\end{equation}
where we have replaced in (\ref{ps para}) $-a_2 z$ by 
$a_3+a_1 x^2$. We now have an expression that depends only
 on two variables~: $x$ and $y$. The problem of determining
 the coefficients $a_1$, $a_2$ and $a_3$ in order for this
 expression to have the same sign for arbitrary values of x
 and y is considerably simpler than the analogous problem
 in the three variables $x$,$y$ and $z$ that must be
 solved in the method of the Lyapunov functionals.

\begin{figure}[htbp]
$$
\epsfxsize=5cm
\epsfbox{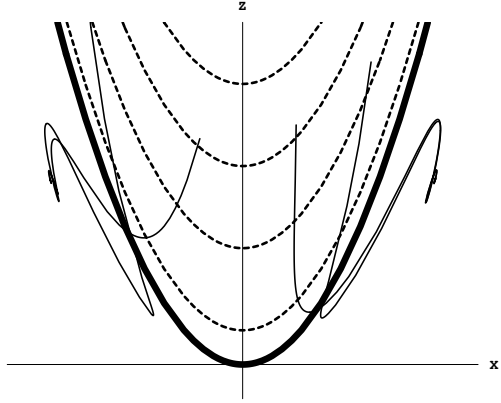}
$$
\caption{The family of semipermeable surfaces (\ref{para})
 for the case $b>2 \sigma$. The bolded surface is the {\it last}
 surface of the family. Some trajectories of the system
 are also shown. The critical points C+ and C-
 are below these surfaces.}
\label{fig para h}
\end{figure}

\begin{figure}[hbtp]
$$
\epsfxsize=5cm
\epsfbox{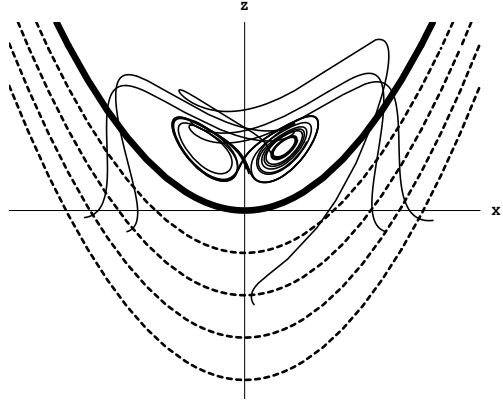}
$$
\caption{The family of semipermeable surfaces(\ref{para})
for the case $b<2 \sigma$. The bolded surface is the {\it last}
surface of the family. Some trajectories of the system are also shown.
The chaotic attractor is above these surfaces.}
\label{fig para b}
\end{figure}

To keep the same sign in (\ref{ps para on}) for arbitrary
values of $x$ and $y$, we must take $a_2=-2 \sigma a_1$. Then
we have~:
\begin{equation}
\vec{N} \cdot \vec{T}/S=a_1(b-2 \sigma)x^2+b a_3 \label{ps para fin}
\end{equation}
As $a_1$ must be nonzero, we can take $a_1=1$ without loss of
 generality.
We now have two different cases~:
\begin{itemize}
\item [i)] $b>2\sigma $, we must take $a_3>0$ in 
order to have a family of semipermeable surfaces.
We show this family, as well as
some trajectories of the system, in fig. \ref{fig para h}.
\item [ ii)] $b<2\sigma $, we must take $a_3<0$ in 
order to have a family of semipermeable surfaces.
We show this family, as well as
some trajectories of the system, in fig. \ref{fig para b}.
\end{itemize}
As we can see from the figures above, in the region filled by
 the surfaces the dynamic of the system is very simple.
The {\it complex} behaviour can only occur in the region of
 phase space that is not occupied by these surfaces.
 In case ii), the global attractor of the system must be located in the 
region $z>2 \sigma x^2$. For the case i), because of the
 presence of the semipermeable surfaces, the flow cannot
 enter the $z>2 \sigma x^2$ region upward and hence the
 homoclinic trajectory cannot exist.

Therefore, motivated by the existence of the first
 integral {\bf a}, valid in the case $b=2\sigma$, we have found
 a family of semipermeable surfaces for arbitrary values 
of the parameters of the system.

As we shall see, new families of semipermeable surfaces can be
 found by using the other integrals of motion.\\

From the case {\bf b}, we deduce that the surfaces $y^2+z^2=k^2$ are
 semipermeable for arbitrary values of k, when $b=1$ and
 $r=0$. Guided by this result, we propose in the general
 case a family of surfaces of the form~:
\begin{equation}
S=a_1(y-c_1)^2+a_2(z-c_2)^2-1=0
\label{cylindre}
\end{equation}
The scalar product $\vec{N} \cdot \vec{T}$ is given by~:
\begin{eqnarray}
\vec{N} \cdot \vec{T} & = &2 (  
 (a_2-a_1)xyz-a_2 b z^2-a_1 y^2+(a_1 r-a_2 c_2)xy+
a_1 c_1 xz  \nonumber \\
& &+ a_2 c_2 b z+a_1 c_1 y-a_1 c_1 r x )  \label{ps cylindre}
\end{eqnarray}
The evaluation of $\vec{N} \cdot \vec{T}$ on the surface
(\ref{cylindre}) becomes far simpler if we take 
$a_2=a_1$, $c_1=0$ and $c_2=r$. After this, (\ref{cylindre})
and (\ref{ps cylindre}) respectively become~:
\begin{equation}
S=a_1(y^2+(z-r)^2)-1=0
\end{equation}
\begin{equation}
\vec{N} \cdot \vec{T}=-2a_1(b z^2+y^2-rbz)
\end{equation}
The scalar product $\vec{N} \cdot \vec{T}$ calculated on 
the surface $S$ is given by~:
\begin{equation}
\vec{N} \cdot \vec{T}/S=
a_1 \left( (1-b)z^2+r(b-2)z+r^2-\frac{1}{a_1} \right)
\label{ps cylindre on}
\end {equation}

\begin{figure}[htbp]
$$
\epsfxsize=5cm
\epsfbox{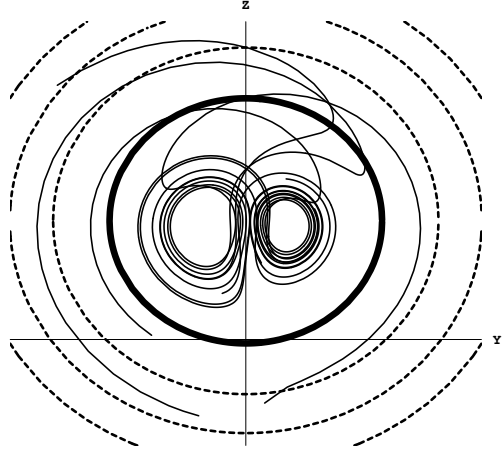}
$$
\caption{Chaotic attractor stuck inside semipermeable cylinders.
The bolded circle corresponds to the {\it smallest} cylinder.}
\label{fig cyl b}
\end{figure}

Note that $\vec{N} \cdot \vec{T}/S$ is a function of
only one variable, as it is the case for expression
(\ref{ps para fin}) for the semipermeable parabolas.
In this case, the surface $S$ is not infinite in the $y$
 and $z$ directions. In particular, the coordinate $z$ 
varies in the interval~:
$r-\frac{1}{\sqrt{a_1}} \leq z \leq r+\frac{1}{\sqrt{a_1}}$.
 In consequence, the quadratic polynomial 
(\ref {ps cylindre on}) must have the same sign only in
 this interval and not for arbitrary values of
 $z$. This condition determines the possible values
 of $a_1$, that can be found by applying the Sturm's theorem. 
The results are as follows~:
\begin{eqnarray}
b\leq 2 & \mbox{,   }& \frac{1}{a_1}\geq r^2 \nonumber \\ 
b\geq 2 & \mbox{,   }&  \frac{1}{a_1}\geq 
\frac{r^2 b^2}{4(b-1)} \label {cylindre min}
\end{eqnarray}
Hence, for arbitrary values of the parameters of the
 system, we have found a family of semipermeable infinite
 cylinders. The radius of these cylinders varies between
 $+ \infty$ and the minimal values given in
 (\ref{cylindre min}). The behaviour of some orbits with
 respect to this family of surfaces is shown in
 fig. \ref{fig cyl b}. The global attractor is
 contained in the region not occupied by these surfaces.\\

\begin{figure}[htbp]
$$
\epsfxsize=5cm
\epsfbox{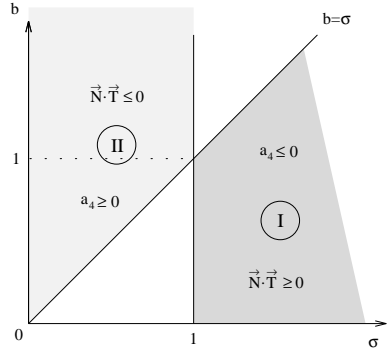}
$$
\caption{Parameters $\sigma$ and $b$ for which surfaces (\ref{cone})
are semipermeable.}
\label{fig hyper}
\end{figure}

\begin{figure}[htbp]
$$
\epsfxsize=5cm
\epsfbox{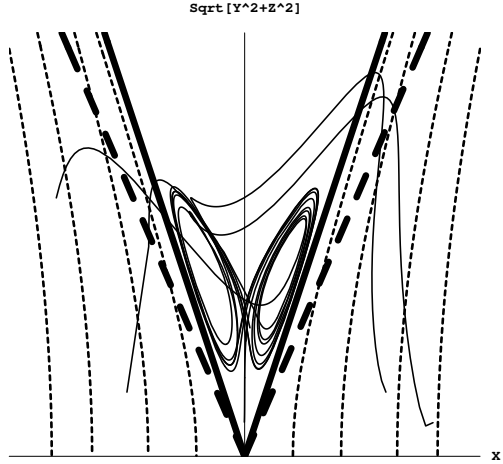}
$$
\caption{Lorenz attractor squeezed between semipermeable
hyperboloids (\ref{cone}). The bolded lines are the last
repelling cones. The thick dashed
 line is the former bound obtained by Doering \& al.
 \cite{doering et gibbon}. The parameters $\sigma$ and $b$
are in region I.}
\label{fig cone b}
\end{figure}

From the integral of motion {\bf\ c}, we deduce the existence of
 the family of semipermeable surfaces $z^2+y^2-rx^2-k=0$, 
with $k$ arbitrary and $b=1$, $\sigma=1$.\\ Guided by this
 result, we propose in the general case the family of surfaces~:
\begin{equation}
S=a_1 x^2+a_2 y^2+a_3 z^2-a_4=0 \; \mbox{ , with }
 a_1 \, a_2 \, a_3 <0 \nonumber
\end{equation}
 The scalar product $ \vec{N} \cdot \vec{T}$ is given by~:
\begin{equation}
\vec{N} \cdot \vec{T}=2 \left(
(a_3-a_2)xyz-a_3b z^2-a_2 y^2-a_1 \sigma x^2+(a_1 \sigma +a_2 r)xy
\right)
\end{equation}
In order to have the same sign for $\vec{N} \cdot \vec{T}$ on 
the surface $S$, we take $a_3=a_2=1$. Then we have~:
\begin{equation}
S=a_1 x^2+ y^2+ z^2-a_4=0 \; \mbox{ , with } a_1<0 \label{cone}
\end{equation}
\begin{equation}
\vec{N} \cdot \vec{T}=2 \left(
-(y^2+b z^2)-a_1 \sigma x^2+(a_1 \sigma+r)xy
\right)
\end{equation}
This expression, calculated on the surface (\ref{cone}),
gives~:
\begin{equation}
\vec{N} \cdot \vec{T}/S=2 \left(
a_1(b- \sigma )x^2+(a_1 \sigma+r)xy+(b-1)y^2-b a_4
\right) \label {ps cone on}
\end{equation}
On the surface (\ref{cone}) the variables $x$ and $y$ vary
 in such a way that the following inequality must be
 satisfied~: $a_1 x^2+y^2 \leq a_4$. Therefore
 (\ref{ps cone on}) must have the same sign for all values
 of $x$ and $y$ that satisfy this inequality. The
 solution of this algebraic problem is not simple. Hence
 we do not give here the technical details of the
 calculations.\\ This condition determines the possible
 values of the coefficient $a_1$~:
\begin{eqnarray}
\frac{-2 \sigma r -(\sigma -1)^2- 
\sqrt{(\sigma -1)^4+4 \sigma r (\sigma -1)^2}}
{2 \sigma ^2}
\leq a_1 \leq \nonumber \\
\frac{-2 \sigma r -(\sigma -1)^2+
\sqrt{(\sigma -1)^4+4 \sigma r (\sigma -1)^2}}
{2 \sigma ^2}
\end{eqnarray}
or
\begin{eqnarray}
\frac{2(b-1)(b-\sigma)-\sigma r 
-2 \sqrt{(b-1)^2(b-\sigma)^2-\sigma r (b-1)(b- \sigma)}}
{\sigma ^2}
\leq a_1 \leq \nonumber \\
\frac{2(b-1)(b-\sigma)-\sigma r 
+2 \sqrt{(b-1)^2(b-\sigma)^2-\sigma r (b-1)(b- \sigma)}}
{\sigma ^2}
\end{eqnarray}
The parameter $a_4$ is arbitrary, and the condition $a_1 < 0$ restricts the 
possible values of the parameters $b$ and $\sigma$, which
 are given in fig. \ref{fig hyper}.
The canonical values $r=28$, $\sigma=10$ and 
$b=\frac{8}{3}$ lie in the region I. For parameters $\sigma$
and $b$ in region I of fig. \ref{fig hyper}, there are
 two zones in phase space 
that are filled with surfaces of the family. The behaviour 
of some trajectories of the system, in relation to the
 semipermeable surfaces is shown in fig. \ref{fig cone b}.

 These results are more restrictive (they give a more precise
 information about the location of the global attractor)
 than similar results obtained recently in \cite {doering et
 gibbon} by employing the method of Lyapunov functions.
 Therefore, our method can locate more
 accurately the global attractor of the system in phase
 space than before.

\begin{figure}[htbp]
$$
\epsfxsize=5cm
\epsfbox{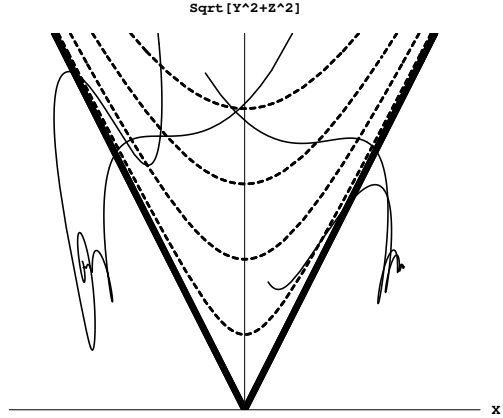}
$$
\caption{Critical points C- and C+ separated by semipermeable
 surfaces (\ref{cone}). The left (resp. right) zone is a part of the bassin
of attraction of C- (resp. C+). The parameters $\sigma$ and $b$
 are in region II.}
\label{fig cone h}
\end{figure}

 In region II of fig. \ref{fig hyper},
the scalar product $\vec{N} \cdot \vec{T}/S$ has opposite
 sign with respect to region I and, in phase space, there are
 two zones that are not filled by the surfaces of the
 family. These two zones are not connected between them
 and the critical points C- and C+ are contained in each
 of the different regions. The behaviour of some
 trajectories of the flow with respect to the
 semipermeable surfaces and the position of the critical
 points C- and C+ are shown in the fig. \ref{fig cone h}.
 If one trajectory enters one of the two free regions in the
 phase space, it cannot exit from it and hence
 cannot pass in the other one. This restriction on the
 behaviour of the orbits prevents the possibility of a
 chaotic behaviour. The trajectories that evolve around
 one of the critical points cannot go to the other free
 region for evolving around the other critical point (we
 make reference to the critical points C- and C+ only).
 It is clear that the homoclinic bifurcation that precedes
 the birth of the chaotic behaviour cannot occur in
 region II of the parameter space.\\
There is still another result for region II of the parameter 
space~: since the flow, once it has entered one of the two free
 regions of the phase space, cannot escape from it, 
it can only go to the critical point C- or C+ lying in this
 region. So each one of the two free zones in
 the phase space is a part of the attraction's basin of
 C- or C+.\\

By applying the same method and guided by the
 form of the integral of motion {\bf d}, we have found a
 new family of
 semipermeable surfaces. We do not give here the technical
 details of the calculations. They are a little more
 complicated than the calculations involved in the previous cases.
 The results are as follows. The family of surfaces~:
 $S=c_1+c_2 x^2-\frac{1}{4 \sigma} x^4+(2 \sigma-b) x y +
\sigma y^2+x^2 z=0 $ are semipermeable in the two following
cases~:
\begin{itemize}
\item[i)] $ b<6 \sigma-2$, $r<2 \sigma-1$, $c_1 \geq 0$ 
and $c_2 \in I_{c_2}$, where $I_{c_2}$ is the interval defined
 by the two real roots of the quadratic polynomial in $c_2$~:
$4 \sigma^2  c_2^2+ 4 \sigma(b+ 2\sigma-b \sigma+2 r \sigma -
6 \sigma^2)c_2+b^2-4 b \sigma - 6 b^2 \sigma - 4 b r \sigma -
4 b^2 r \sigma + 4 \sigma ^2 +24 b \sigma ^2 +
9 b^2 \sigma ^2 + 8 r \sigma ^2 + 20 b r \sigma ^2 +
 4 r^2 \sigma ^2 -24 \sigma^3-36 b \sigma^3 -
24 r \sigma^3 +36 \sigma ^4$. In this case, the sign of the
 scalar product $\vec{N} \cdot \vec{T}/S$ is positive. The
 canonical values of the Lorenz's parameters do not satisfy the
 above two inequalities between $r$, $b$ and $\sigma$.
\item [ii)]  $ b>6 \sigma-2$, $r>2 \sigma-1$, $c_1=0$ 
and $c_2 \in I_{c_2}$.\\ In this case, the sign of
 $\vec{N} \cdot \vec{T}/S$ is negative. This family of surfaces
 divides the phase space in three regions. Only one of them is
 filled by the surfaces of the family. The two free regions are
 disconnected and the critical points C+ and C- are located
 in different regions. Here, as in one of the cases analysed above,
 the homoclinic bifurcation, and hence the chaotic behaviour,
 is not possible. The
 critical points C+ and C- are stable and each of the two
 free regions is part of the basin of attraction of each critical
 point.
\end{itemize}

Finally, guided by the form of the first integral {\bf e}, we
 have found another family of semipermeable surfaces.
Let us consider the family of surfaces~:
\begin{equation}
S=c_1+\frac{2-3b+b^2-c_2+2 \sigma -b \sigma -2 r \sigma}{2
 \sigma}x^2-\frac{1}{4 \sigma}x^4+(2-b)xy+\sigma y^2+
(c_2+x^2)z=0 \label{fact4}
\end{equation}
These surfaces are semipermeable in four different cases~:
\begin{itemize}
\item [$\alpha$ ] $b>2(\sigma+1)$, $\sigma<1$, $c_1  c_2 \leq 0$,
 $c_1<\frac{(c_2+k_3)(c_2k_2+b^2k_3)}
{64(\sigma-1)\sigma(\sigma+1)}$; \\ here
 the flow is crosssing the surfaces downward.
\item [$\beta$ ] $b>2(\sigma+1)$, $\sigma<1$, $c_1  c_2
 \leq 0$, $c_2>\frac{-bk_3}{b-4 \sigma +4}$, 
$c_1<\frac{2-3b+b^2-c_2+2 \sigma-b \sigma-2 r \sigma}
{2 \sigma}$; \\ here
 the flow is crossing the surfaces downward too.
\item [$\gamma$ ] $b<2(\sigma+1)$, $\sigma>1$, $c_1  c_2 \geq 0$,
 $c_1>\frac{(c_2+k_3)(c_2k_2+b^2k_3)}
{64(\sigma-1)\sigma(\sigma+1)}$; \\ here
 the flow is crossing the surfaces upward.
\item [$\delta$ ] $b<2(\sigma+1)$, $\sigma>1$, $c_1  c_2 \geq 0$,
 $(b-4 \sigma+4)c_2<-bk_3$\\ $c_1>\frac{2-3b+b^2-c_2+2
\sigma-b \sigma-2 r \sigma}{2 \sigma}$; \\ here
 the flow is crossing the surfaces upward too.
\item [] The quantities $k_2$ and $k_3$ are given by~:
$k_2=b^2+8b(\sigma-1)+16(1-\sigma^2)$ and
$k_3=2b-b^2-4 \sigma+2 b \sigma+4 r \sigma$.
\end{itemize}
\begin{figure}[htbp]
$$
\epsfxsize=5cm
\epsfbox{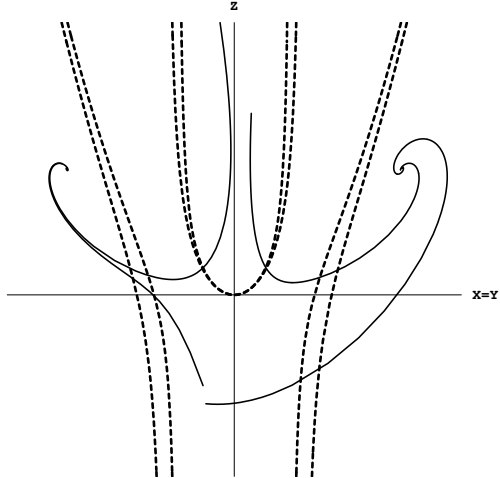}
$$
\caption{Critical points separated by semipermeable
 surfaces of type (\ref{fact4}) in case $\alpha$.
The far left (resp. right) zone is a part of the bassin of
 attraction of C- (resp. C+).}
\label{fig fact4 h}
\end{figure}
\begin{figure}[htbp]
$$
\epsfxsize=5cm
\epsfbox{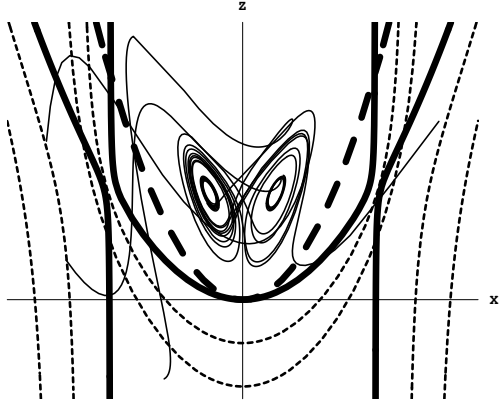}
$$
\caption{Lorenz attractor enclosed by semipermeable surfaces
 of type (\ref{fact4}) in case $\gamma$. The form of these
 surfaces for $x^2<-c_2$
is a sink, what we cannot see in the figure, which is a projection.
The bolded curve is given by the equality in expression
 (\ref{uppest}). The thick dashed curved
is the {\it last} parabola.}
\label{fig fact4 b}
\end{figure}
If $c_2>0$, each surface of the family is connected;
 if $c_2<0$, all the surfaces are disconnected~: they are
 divided in three parts.

In case $\alpha$, if we  take
 $r>\frac{1}{4 \sigma} (b-2)(b-2 \sigma)$, $c_2 \in
 [-b^2 \frac{k_3}{k_2} \mbox{;} -k_3]<0$ and then $c_1>0$,
we then have disconnected semipermeable surfaces and the
 critical points are under the
 surfaces and separated by them. This is another configuration
 where we know a part of the basin of attraction of each
 critical point and where the homoclinic trajectory
cannot exists (see fig. \ref{fig fact4 h}).

The case $\gamma$ gives us information about the space extension
 of the chaotic attractor ($r=28$, $\sigma=10$, $b=\frac{8}{3}$).
 In this case, when $c_2<0$ and $c_1 \geq 0$, the surfaces 
are disconnected and the flow crosses them upward
 (see fig. \ref{fig fact4 b}).
 The uppermost surface (for $c_1=0$ and $c_2=-k_3$)
is an additional bound for the Lorenz attractor. Hence it lies
 entirely in the zone of the phase space where~:
\begin{equation}
{\large z \geq - \frac{\frac{2-3b+b^2-c_2+2 \sigma -b \sigma -2 r \sigma}
{2 \sigma}x^2-\frac{1}{4 \sigma}x^4+(2-b)xy+\sigma y^2}
{x^2-k_3} } \label {uppest}
\end{equation}
(for Lorenz's canonical values $k_3=1131.56$).\\

\begin{figure}[htbp]
$$
\epsfxsize=6cm
\epsffile{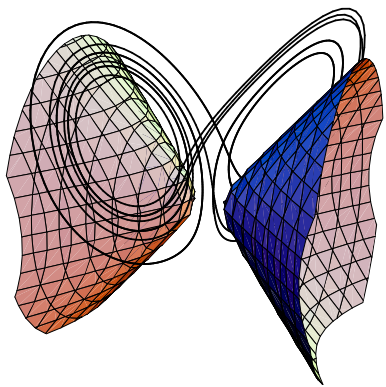}
$$
\caption{Lorenz's attractor and a semipermeable surface
of the family (\ref{ellipsoid}).
When we consider all the surfaces of the family, stronger
bounds on the location of the attractor are obtained.}
\label{fig cone z c3}
\end{figure}

We have also found several semipermeable families of
 ellipsoids. In fact, we have generalised results given in
 \cite{doering et gibbon,sparrow,jackson}.
Surfaces like
\begin{equation}
S=\frac{c_3-r}{\sigma} x^2+y^2+(z-c_3)^2=R \label {ellipsoid}
\end{equation}
are semipermeable for the following cases~:
\begin{itemize}
\item $c_3 > r$ for arbitrary $\sigma$, $b$, $r$ and for values
 of $R$ as in fig. \ref{fig elli}.
\begin{figure}[htbp]
$$
\epsfxsize=5cm
\epsfbox{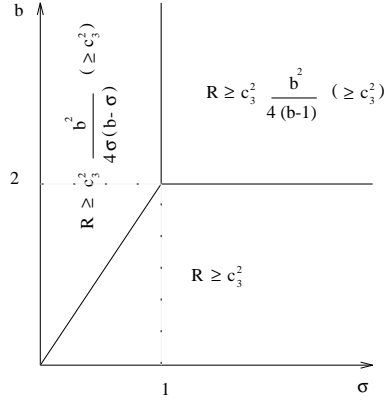}
$$
\caption{Parameters for which surfaces (\ref{ellipsoid}) with $c_3 >r$
 are semipermeable.}
\label{fig elli}
\end{figure}	
The interest of having one free parameter (here $c_3$) in
 addition to $R$ is that we may
 lower the ellipsoids in phase space with $c_3$ and so restrict
 more tightly
 the region in which the chaotic attractor lies (with considering
 the envelope of all the smallest (with R minimum) ellipsoids
when $c_3$ varies in $]r \mbox{; } + \infty[$). These results contain
known results about ellipsoids and several new ones.
\item if $c_3=r$ then $S$ is the cylinder
 which we have studied above.
\item if $c_3<r$ then $S$ is an hyperboloid of
revolution. The revolution axis is $\{z=c_3,y=0\}$.
This surface is semipermeable for
 $ b< \sigma \mbox{ and } \sigma >1 $
 (a case which includes Lorenz's canonical values)
 and $ R \leq \frac{b^2}{4 \sigma(b- \sigma)} {c_3}^2 $
For this case the scalar product is positive $\forall
 \; \{x,y\} \in S $.
The flow crosses the last surface $\, \frac{c_3-r}
{\sigma}x^2+y^2+(z-c_3)^2=\frac{b^2}{4 \sigma(b-
 \sigma)} {c_3}^2 $ 
outwards. This new information sharpens the bounding frontiers
 of the attractor as we can see in fig. \ref{fig cone z c3}.
\end{itemize}

Sparrow \cite{sparrow} has conjectured that all
 trajectories of the Lorenz system eventually enter and
 remain in the region $z \geq 0$ for all parameters values
 $r$, $\sigma$ and $b$ (note that the plane $z=0$ is not
 semipermeable). Sparrow proved this conjecture for the
 case $b \leq \sigma +1$ by using the method of Lyapunov 
functions.

The existence of semipermeable parabolas $z=\frac{x^2}
{2 \sigma}+a_0$ for
$a_0<0$ and $b<2 \sigma$ also proves the conjecture, but
 for different values of parameters $\sigma$ and b. Indeed,
 in this case the flow is crossing all
the parabolas upward. It doesn't mean that the $z=0$
 plane itself is semipermeable, but the flow has to
 end up with crossing upward 
the last parabola $z=\frac{x^2}{2 \sigma}$. The
 parabolas are in fact
driving the flow to the phase space zone where
 $z>\frac{x^2}{2 \sigma} \geq 0$.

The cylinders family~: $y^2+(z-r)^2=R$ with $R \geq r^2$
 and $b \leq 2$ also drives the flow inside the 
{\it smallest}
 cylinder ($R=r^2$) which lies in the $z \geq 0$
phase region.

\begin{figure}[htbp]
$$
\epsffile{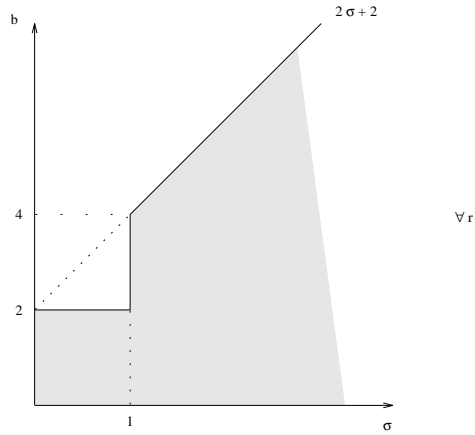}
$$
\caption{Range of parameters $\sigma$ and b for which the
 flow eventually enters 
the phase space zone $z>0$ ($\forall r$)}
\label{fig z plan}
\end{figure}

The surfaces (\ref{fact4}), in case $\gamma$,
tell us also that the flow eventually crosses the uppermost surface
 (given by the equality in (\ref{uppest})) upward. This
 surface lies entirely in the $z \geq 0$ half space
 (for $x^2 < -c_2$). The results of this work expand the
 region of the parameter space for which the flow
 eventually enters the zone of phase space with $z>0$.
This region is shown in  fig. \ref{fig z plan}.\\

In conclusion, inspired by the integrals of motion that
 exist for particular values of the parameters $b$, $\sigma$
 and $r$, we were able to find several families of
 surfaces, all crossed in the same direction by the
 flow associated to the system.

From these results, we have deduced a rich quantity of 
information about the geometrical location of the global
 attractor of the system. This information is more restrictive
 than similar results that had been found by the method
 of Lyapunov functions. When compared to the Lyapunov technique,
we see that the fundamental advantage of this new method
is that one now has to study functions with one less variable.

Moreover, we have obtained information
 about the spread of the basin of attraction of the
 critical points C- and C+ when they are stable. We have
 also determined regions of the parameter space where the
 chaotic behaviour is not possible.

It is clear that the method used in this paper can be
 applied to other 3-d dissipative dynamical systems that
 the Lorenz one. We have choosen the latter owing to
 the great importance that this system has played
 in the study of chaotic dynamics. 

\end{document}